\documentclass[reprint,longbibliography]{revtex4-1} 
\usepackage{amssymb,amsmath}
\usepackage{graphicx,import}
\usepackage{placeins}
\usepackage{hyperref}
\usepackage{color}
\usepackage{longtable}
\usepackage{amsthm}
\usepackage{algorithm}
\usepackage{algpseudocode}
\usepackage{upgreek}
\usepackage[nameinlink,poorman]{cleveref}

\crefname{equation}{Eq.}{Eqs.}
\crefname{figure}{Fig.}{Figs.}
\crefname{table}{Table}{Tables} 
\crefname{section}{Section}{Sections}
\crefname{chapter}{Chapter}{Chapters}
\crefname{appendix}{Appendix}{Appendices}
\crefname{algorithm}{Algorithm}{Algorithms}

\graphicspath{{./paperfigs/}} 

\begin{document}

\title{Autonomous adaptive noise characterization in quantum computers}

\author{Riddhi Swaroop Gupta} 
\email{riddhi.sw@gmail.com}
\affiliation{ARC Centre of Excellence for Engineered Quantum Systems, School of Physics, The University of Sydney, New South Wales 2006, Australia}

\author{Alistair R. Milne} 
\affiliation{ARC Centre of Excellence for Engineered Quantum Systems, School of Physics, The University of Sydney, New South Wales 2006, Australia}

\author{Claire L. Edmunds} 
\affiliation{ARC Centre of Excellence for Engineered Quantum Systems, School of Physics, The University of Sydney, New South Wales 2006, Australia}

\author{Cornelius Hempel} 
\affiliation{ARC Centre of Excellence for Engineered Quantum Systems, School of Physics, The University of Sydney, New South Wales 2006, Australia}

\author{Michael J. Biercuk}
\affiliation{ARC Centre of Excellence for Engineered Quantum Systems, School of Physics, The University of Sydney, New South Wales 2006, Australia}

\begin{abstract} New quantum computing architectures consider integrating qubits as sensors to provide actionable information useful for decoherence mitigation on neighboring data qubits, but little work has addressed how such schemes may be efficiently implemented in order to maximize information utilization.  Techniques from classical estimation and dynamic control, suitably adapted to the strictures of quantum measurement, provide an opportunity to extract augmented hardware performance through automation of low-level characterization and control.  In this work, we present an autonomous learning framework, Noise Mapping for Quantum Architectures (NMQA), for adaptive scheduling of sensor-qubit measurements and efficient spatial noise mapping (prior to actuation) across device architectures. Via a two-layer particle filter, NMQA receives binary measurements and determines regions within the architecture that share common noise processes;  an adaptive controller then schedules future measurements to reduce map uncertainty.  Numerical analysis and experiments on an array of trapped ytterbium ions demonstrate that NMQA outperforms brute-force mapping by up-to $18$x ($3$x) in simulations (experiments), calculated as a reduction in the number of measurements required to map a spatially inhomogeneous magnetic field with a target error metric. As an early adaptation of robotic control to quantum devices, this work opens up exciting new avenues in quantum computer science.
\end{abstract}

\maketitle


\section{Introduction}
In the NISQ era~\cite{Preskill2018quantumcomputingin}, the impacts of decoherence and hardware error remain dominant considerations in the drive to achieve quantum advantages using realistic multi-qubit architectures~\cite{yao2012scalable,monroe2014large,veldhorst2017silicon,jones2012layered,kielpinski2002architecture,franke2019rent}. Prior to the deployment of full quantum error correction, control solutions implemented as a form of ``quantum firmware''~\cite{BallPRApplied2016} at the lowest level of the quantum computing software stack~\cite{jones2012layered} provide an opportunity to improve hardware error rates using both open-loop dynamic error suppression~\cite{Brown2004,True, Khodjasteh2009dcg,SoareNatPhys2014,ViolaFFF,Kabytayev2014} and closed-loop feedback stabilization.  There is substantial opportunity for autonomous optimization routines to be built and deployed in the drive for improved hardware performance ~\cite{venturelli2018compiling,murali2019noise,shi2019optimized,venturelli2018optimization,tannu2018case}, and such approaches increase in importance as quantum computer system sizes grow.

Closed-loop stabilization is a common low-level control techniques for classical hardware stabilization \cite{gelb1974applied,landau2011adaptive}, but its mapping to quantum devices faces challenges in the context of quantum state collapse under projective measurement~\cite{Mavadia2017,gupta2018machine}.  This can be partially addressed through the adoption of an architectural approach embedding additional qubits as sensors at the physical level to provide actionable information on decoherence mechanisms, {\em e.g.} real-time measurements of ambient field fluctuations~\cite{Cappellaro2014}.  The objective is to spatially multiplex the complementary tasks of noise sensing and quantum data processing in a multi-qubit architecture~\cite{BrownArch}.  Fundamental to such an approach is the existence of spatial correlations in noise fields, permitting information gained from a measurement on a so-called spectator qubit to be deployed in stabilizing proximal data qubits~\cite{Cappellaro2016} used in the computation. 

Making the spectator-qubit paradigm practically useful - even in medium-scale architectures - requires spatial mapping of the underlying noise fields~\cite{postler2018experimental} in order to determine which qubits may be actuated upon using information from a specific sensor. This is because spatial inhomogeneities in background fields can cause decorrelation between spectator and data qubits such that feedback stabilization becomes ineffective or even detrimental. This mapping procedure thus becomes one of the first steps prior to designing and deploying more sophisticated control routines for hardware stabilization.  As noted above, stabilizing hardware through the quantum firmware layer provides both an opportunity and a need to leverage high-efficiency autonomous algorithms, including in device characterization and noise mapping.

In this manuscript, we introduce a new framework for autonomous learning, denoted Noise Mapping for Quantum Architectures (NMQA), to efficiently build a map of unknown spatial fields in a multi-qubit quantum computer. NMQA is a classical filtering algorithm operated at the quantum firmware level, and is specifically designed to accommodate the nonlinear, discretized measurement model associated with projective measurements on individual spectator qubits.  The algorithm autonomously schedules measurements across a multi-qubit device and shares classical state information between qubits to enable noise mapping and the discovery of effective noise correlation lengths with enhanced efficiency relative to brute force approaches.  We implement the NMQA framework via a novel two-layer particle filter and an autonomous real-time controller. Our algorithm iteratively builds a map of the noise field across a multi-qubit architecture in real-time by maximizing the information utility obtained from each physical measurement. This, in turn, enables the controller to adaptively determine the highest-value measurement to be performed in the following step.  We evaluate the performance of this algorithm in test cases by both numeric simulation on 1D and 2D qubit arrays, and application to real experimental data derived from Ramsey measurements on a 1D crystal of trapped ions. Our results demonstrate that NMQA outperforms brute-force measurement strategies by a reduction of up to $18\times$ ($3\times$) in the number of measurements required to estimate a noise map with a target error for simulations (experiments).  These results hold for both 1D and 2D qubit regular arrays subject to different noise fields.   

\section{The NMQA Framework}

We consider a spatial arrangement of $d$ qubits as determined by a particular choice of hardware. An unknown, classical field exhibiting spatial correlations extends over all qubits on our notional device, and corresponds to ambient noise fields in multi-qubit operating architectures.  Our objective is to build a map of the underlying spatial variation of the noise field with the fewest possible single-qubit measurements, performed sequentially. We conceive that each measurement is a \mbox{single-shot} Ramsey-like experiment in which the presence of the unknown field results in a measurable phase shift between a qubit's basis states at the end of a fixed interrogation period. This phase is not observed directly, but rather inferred from data, as it parameterizes the Born probability of observing a ``0'' or ``1'' outcome in a projective measurement on the qubit; our algorithms take these discretized binary measurement results as input data. The desired output at any given iteration $t$ is a map of the noise field, denoted as a set of unknown qubit phases, $F_t$, inferred from the binary measurement record up to $t$.  

\begin{figure}
	\includegraphics[scale=1]{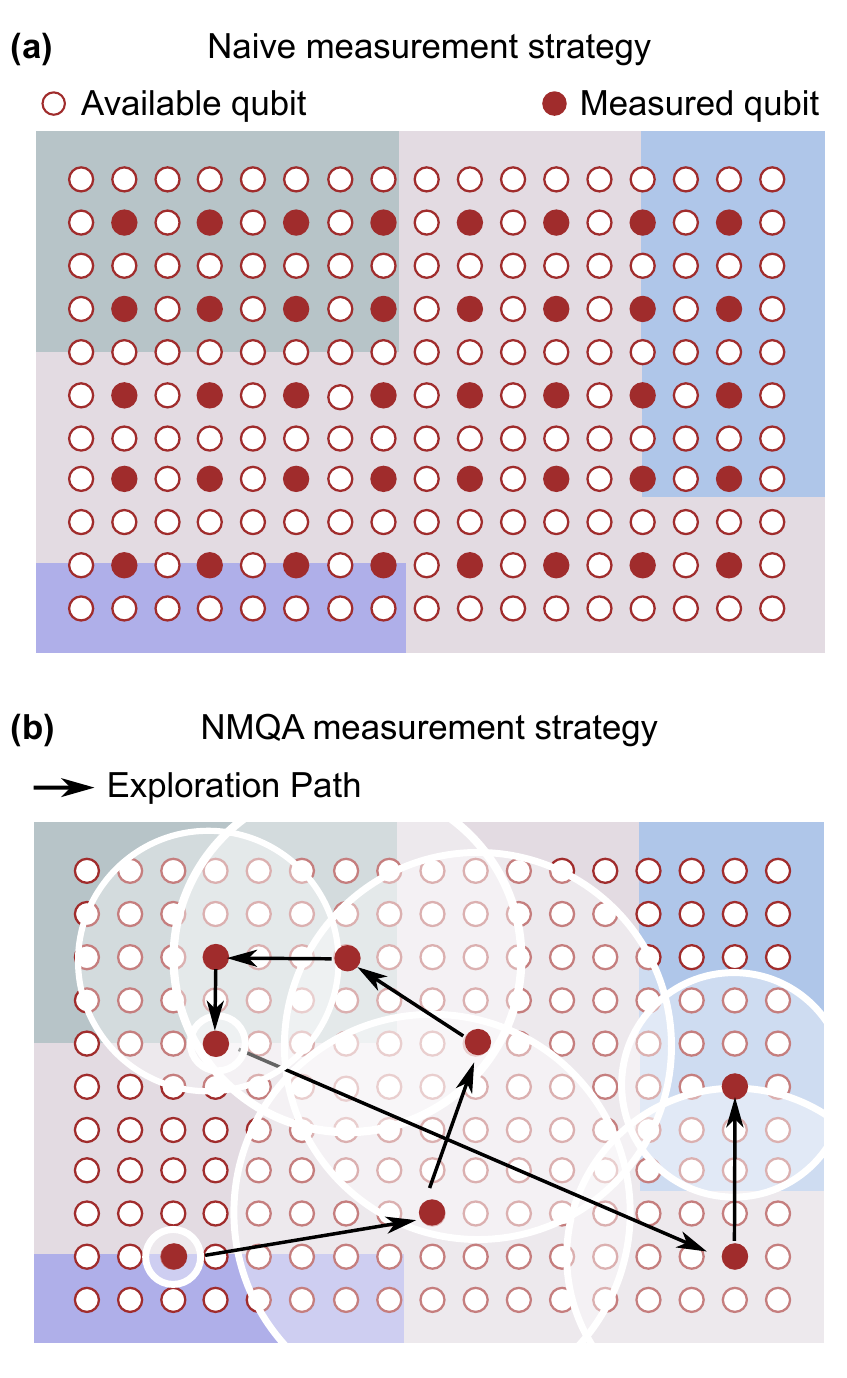}
	\caption{\label{fig_paper_overview} Difference between a naive brute force and the NMQA measurement strategy in reconstructing an inhomogenous background field. A spatial arrangement of qubits (red circles) is shown with a true unknown field with finite correlations (colored regions). (a) The naive strategy measures the field across the array using a regular grid (red filled circles). (b) The NMQA strategy iteratively chooses which qubit to measure next (black arrows) and additionally stores state estimation information in a form shared across local neighborhoods (white shaded circles), which reflects the spatial characteristics of the underlying map.}  	
\end{figure} 

The simplest approach to the mapping problem is to undertake a brute-force, ``naive'' strategy in which one uniformly measures sensor-qubits across the array, depicted schematically in ~\cref{fig_paper_overview}(a). By extensively sampling qubit locations in space repeatedly, one can build a map of the underlying noise fields through the collection of large amounts of measurement information. Evidently, the naive, brute-force measurement approach is highly inefficient as it fails to recognize and exploit spatial correlations in the underlying noise field that may exist over a length-scale that exceeds the inter-qubit spacing. This is a particular concern in circumstances where qubit measurements are a costly resource {\em e.g.} in time, classical communications bandwidth, or qubit utilization when sensors are dynamically allocated. 

The NMQA framework shown in \cref{fig_paper_overview}(b) stands in stark contrast, taking inspiration from the problem of simultaneous localization and mapping in robotic control~\cite{cadena2016past,bergman1999recursive,stachniss2014particle,durrant2006simultaneous,bailey2006simultaneous,murphy2000bayesian,howard2006multi,thrun2005probabilistic,thrun1998probabilistic}. Here the underlying spatial correlations of the noise are mapped using a learning procedure, which adaptively determines the location of the next, most-relevant measurement to be performed based on past observations. With this approach, NMQA reduces the overall resources required for spatial noise mapping by actively exploiting the spatial correlations to decide whether measurements on specific qubits provide an information gain.

We approximate the theoretical non-linear filtering problem embodied by NMQA using particle-filtering techniques. Particle filters are used to propagate and update classical probability distributions in cases where they are expected to undergo highly non-linear transformations
(cf. standard particle filters \cite{doucet2001introduction} and their use in classical robotic mapping \cite{beevers2007fixed,grisettiyz2005improving,poterjoy2016localized}). In our application, the choice of a particle filter to implement core NMQA functionality accommodates the non-linear measurement model associated with the discretized outcomes given by projective qubit readout. 

\begin{figure*}[t!]
	\includegraphics[scale=0.58]{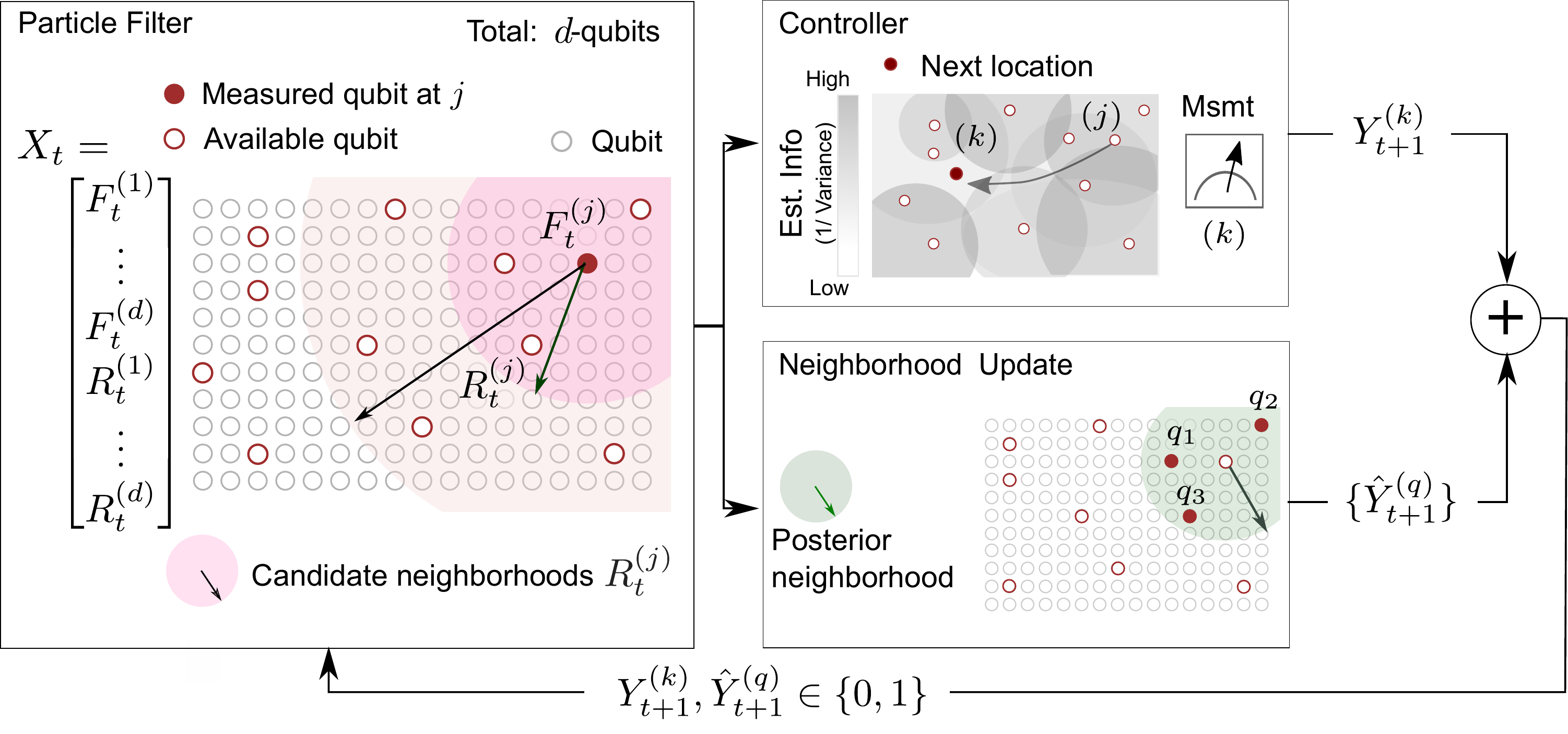}
	\caption{\label{fig_intro_highlevel_v2} Schematic overview of a NMQA iteration. A particle filter estimates the map, $F_t$, and discovers neighborhoods (circular shaded) parameterized by $R_t^{(j)}$ for sharing state information about site $j$. Posterior state estimates from the particle filter are used by the controller, to choose site $k$ as the location of the next physical measurement, $Y^{(k)}_{t+1}$ based on regions of highest estimated uncertainty (middle top). Meanwhile, posterior state estimates are also used to share information at $j$ within the posterior neighborhood, $Q$ via data messages, $\hat{Y}^{(q)}_{t+1}$, between all neighbouring qubits $q \in Q$ (middle bottom) before commencing the next iteration. The input / output information loop is the outermost loop formed by a single physical measurement (in notation, $Y$) and a set of data messages (in notation, $\hat{Y}$). The arrow to site $k$ from $j$ does not relate to the transfer of any information, but rather, adaptive measurement scheduling by the controller to maximize information utility from the next physical measurement choice $k$.}
\end{figure*} 

In our particle filtering implementation, at every iteration $t$, the information about the true map is contained in the state vector, $X_t$. Here the word `state' refers to all quantities being inferred statistically from data, as opposed to an actual physical quantum state. Aside from information about the true map, the state vector, $X_t$, additionally contains information, $R_t$, which approximate spatial gradients on the true map. The probability distribution of $X_t$ conditioned on data is called the posterior distribution and it summarizes our best knowledge of the true map and its approximate spatial gradient information given past measurement outcomes. The posterior distribution is approximately represented as a discrete collection of weighted `particles'. Each particle has two properties: a position and a weight. The position of the particle is effectively a hypothesis about the state $X_t$ i.e. a sample from the probability distribution for $X_t$ given observed data. The weight specifies the likelihood or the importance of the particle in the estimation procedure. All weights are initially equal at step $t=0$ and after receiving measurement data at each step, the particles are `re-sampled'. This means that the original set of particles at step $t$ is replaced by a set of ``offspring'' particles, where the probability that a parent is chosen to represent itself in the next iteration (with replacement) is directly proportional to its weight. Over many iterations, only the particles with the highest weights survive and these surviving particles form the estimate of the distribution of $X_t$ given data, in our algorithm. At any $t$, the estimate of $X_t$ can be obtained as the first moment of the posterior particle distribution, and similarly, true state uncertainty is estimated from the empirical variance of the particle distribution.

The key operating principle of NMQA is that we locally estimate the map value, $F_t^{(j)}$ (a qubit phase shift with value between $[0, \pi]$ radians), before globally sharing the map information at the measured qubit $j$ with neighboring qubits $q \in Q_t$ in the vicinity of $j$. The algorithm is responsible for determining the appropriate size of the neighborhood, $Q_t$, parameterized by the length-scale value, $R_t^{(j)}$ (left panel, \cref{fig_intro_highlevel_v2}). In practice, $Q_t$ eventually represents the set of qubits in a posterior neighborhood at $j$ at the end of iteration $t$, and this set shrinks or grows as inference proceeds. We are ignorant, a priori, of any knowledge of $R_{t=0}^{(j)}$ and estimates take values between $R_{min}$, approximately the inter-qubit spacing, and $R_{max}$, the size of the qubit array, in units of distance. The collection of map values and length-scales, $ X_t^{(j)}:=\{F_t^{(j)}, R_t^{(j)}\} $, at every qubit, $j = 1, 2, \hdots d$, is depicted as an extended state vector.  

We rely on an iterative maximum likelihood procedure within each iteration of the particle filter to solve the NMQA inference problem, as the size of the state-space generally impedes analytic solutions even in classical settings~\cite{thrun2005probabilistic,thrun2001probabilistic}. In each iteration, we update $F_t$ assuming $X_{t-1}$ is known, and subsequently update $R_t$ based on $F_t$. This structure is unique and requires two particle types; $\alpha$ particles carry information about the full state vector, $X_t$, while $\beta$ particles discover the optimal information-sharing neighborhood size, $R_t^{(j)}$, around qubit $j$. The two different types of particle sets are then used to manipulate the joint probability distribution defined over $F_t$ and $R_t$ such that we numerically implement an iterative maximum likelihood procedure. The final result of the particle filtering step in \cref{fig_intro_highlevel_v2} is a posterior estimate of $X_t$ which represents our best knowledge given measurement data.

This estimate is now passed to an autonomous measurement scheduler, the NMQA controller, which attempts to maximize the information utility from each measurement. The NMQA controller adaptively selects the location $k$ for the next physical measurement by choosing the region where posterior state variance is maximally uncertain (\cref{fig_intro_highlevel_v2}, top-middle panel), and that new measurement outcome, once collected, is denoted $Y_{t+1}^{(k)}$ (a posterior state variance is typically estimated using the properties of particles inside the filter \cite{bain2009}). Meanwhile, the posterior state information at step $t$ is also shared with the set $Q_{t}$ in the vicinity of qubit $j$ (\cref{fig_intro_highlevel_v2}, bottom-middle panel). The shared information within this neighborhood is denoted by the set $\{\hat{Y}_{t+1}^{(q)}, q \in Q_{t+1}\}$ and serves as an effective state estimate that is taken as an input to the algorithm in a manner similar to a physical measurement. Jointly, the new actual physical measurement, $Y_{t+1}^{(k)}$, and the set of shared information $\{ \hat{Y}_{t+1}^{(q)} \} $ form the inputs for the next iteration, $t+1$.  Further technical information on the overall NMQA implementation is presented in \emph{Methods}.

We now provide a summary of the particle-filtering implementation used here, and highlight modifications that are unique to NMQA.  Under the conditions articulated above, the NMQA filtering problem requires only two specifications: a prior or initial distribution for the state vector, $X_0$, at $t=0$, and a likelihood function that assigns each particle with an appropriate weight based on measurement data. Assuming a uniform prior, we only need to define the global likelihood function incorporating both particle-types. We label $\alpha$-particles by the set of numbers $\{1, 2, \hdots, n_\alpha\}$; for each $\alpha$ particle, we also associate a set of $\beta$ particles denoted $\beta^{(\alpha)}$, with $\beta$ taking values $\{1, 2, \hdots, n_\beta \}$.  Each $\alpha$ particle is weighted or scored by a so-called likelihood function. This likelihood function is written in notation as $ g_1(\lambda_1, Y_t^{(j)})$, where $\lambda_1$ is a parameter of the NMQA model and $ Y_t^{(j)}$ makes explicit that only one physical measurement at one location is received per iteration. A single $\beta^{(\alpha)}$-particle inherits the state from its $\alpha$-parent, but additionally acquires a single, uniformly distributed sample for $R_t^{(j)}$ from the length-scale prior distribution.  The $\beta$-particles are scored by a separate likelihood function, $g_2(\lambda_2, Q_t)$, where $\lambda_2$ is another parameter of the NMQA model. Then the total likelihood for an $(\alpha, \beta^{(\alpha)})$ pair is given by the product of the $\alpha$ and $\beta$ particle weights. 

The functions $g_1(\lambda_1, Y_t^{(j)})$ and $g_2(\lambda_2, Q_t)$ are likelihood functions used to score particles inside the NMQA particle filter, and their mathematical definitions can be found in the Supplementary Materials. These functions are derived by representing the noise affecting the physical system via probability density functions. The function $g_1(\lambda_1, Y_t^{(j)})$ describes measurement noise on a local projective qubit measurement as a truncated or a quantized Gaussian error model to specify the form of the noise density function.  The function $g_2(\lambda_2, Q_t)$ represents the probability density of ``true errors'' arising from choosing an incorrect set of neighbourhoods while building the map. For each candidate map, the value of the field at the center is extrapolated over the neighborhood using a Gaussian kernel. Thus, what we call ``true errors'' are the differences between the values of the true, spatially continuous map and the NMQA estimate, consisting of a finite set of points on a map and their associated (potentially overlapping) Gaussian neighborhoods. We assume that these errors are truncated Gaussian distributions with non-zero means and variances over many iterations.

The two free parameters, $\lambda_1, \lambda_2 \in [0,1]$, are used to numerically tune the performance of the NMQA particle filter. Practically, $\lambda_1$ controls how shared information is aggregated when estimating the value of the map locally and $\lambda_2$ controls how neighborhoods expand or contract in size with each iteration. Additionally, the numerically tuned values of $\lambda_1, \lambda_2 \in [0,1]$ provide an important test of whether or not the NMQA sharing mechanism is trivial in a given application. Non-zero values suggest that sharing information spatially actually improves performance more than just locally filtering observations for measurement noise. As $\lambda_1, \lambda_2 \to 1$, the set $\{\hat{Y}_{t+1}^{(q)} \}$, is treated as if they were the outcomes of real physical measurements by the algorithm.  However, in the limit $\lambda_1, \lambda_2 \to 0$, no useful information sharing in space occurs, and NMQA effectively reduces to the naive measurement strategy. 

Detailed derivations for mathematical objects and computations associated with NMQA, and an analysis of their properties using standard non-linear filtering theory, will be provided in a forthcoming technical manuscript.

\section{Results \label{sec:results}}

Application of the NMQA algorithm using both numerical simulations and real experimental data demonstrates the capabilities of this routine for a range of spatial arrangements of $d$ qubits in 1D or 2D. Our evaluation procedure begins with the identification of a suitable metric for characterizing mapping performance and efficiency. We choose a Structural SIMilarity Index (SSIM) ~\cite{wang2004image}, frequently used to compare images in machine learning analysis. This metric compares the structural similarity between two images and is defined mathematically in \textit{Methods}.  It is found to be sensitive to improvements in the quality of images while giving robustness against {\em e.g.} large single-pixel errors that frequently plague norm-based metrics for vectorized images~\cite{chen2009similarity,wang2009mean}. In our approach, we compare the true map and its algorithmic reconstruction by calculating the SSIM; a score of zero corresponds to ideal performance, and implies that the true map and its reconstruction are identical. 

\begin{figure}[h]
	\includegraphics[scale=1]{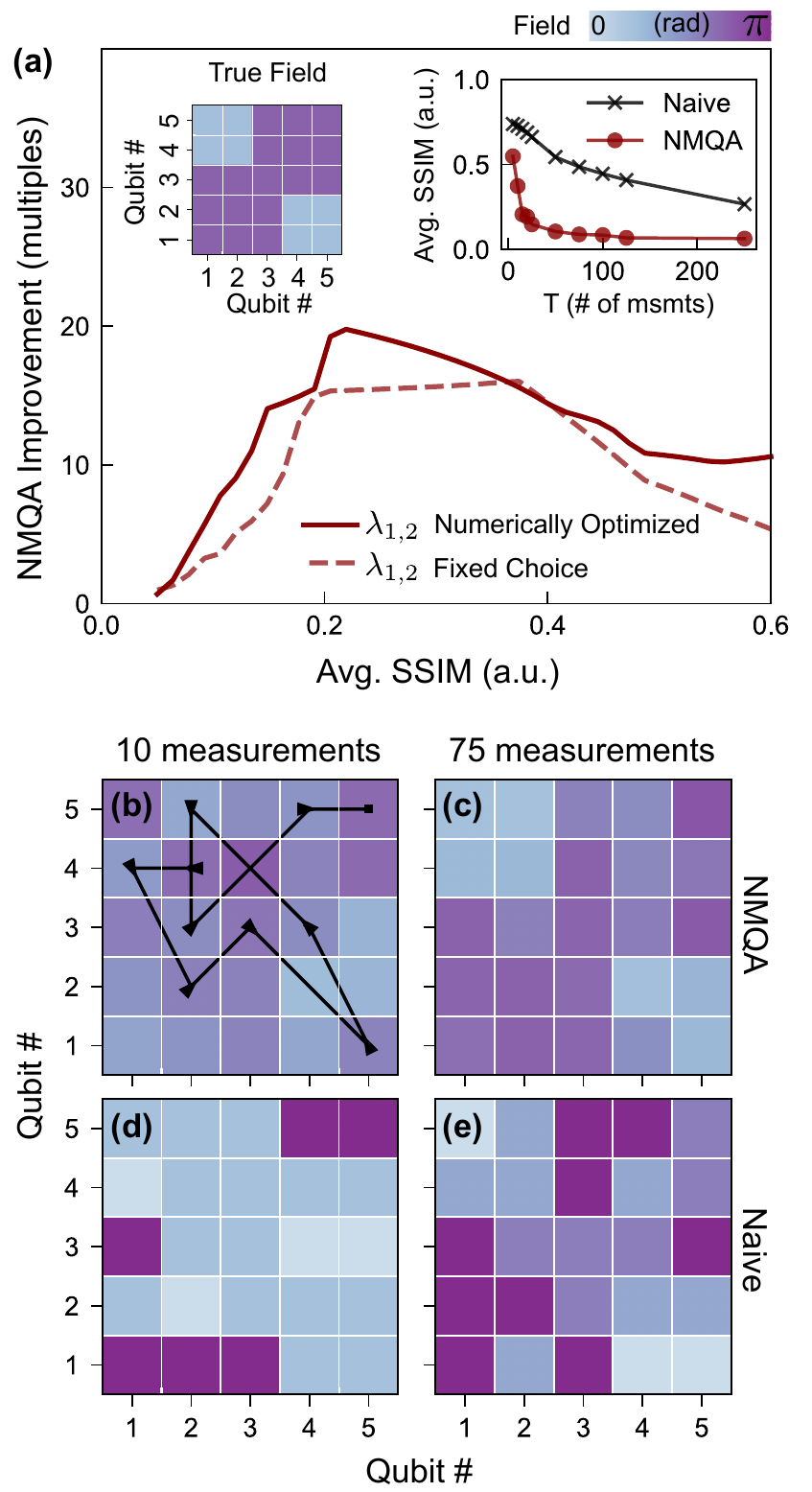}
	\caption{\label{SSIM1_collated} (a) Upper left inset: 2D array of 25 qubits with `square' field shown as a colorscale.  Upper right inset: Avg. SSIM score vs total measurement budget $T$ for NMQA and naive algorithms averaged over 50 trials. Main panel: ratio of naive to NMQA measurements vs. Avg. SSIM score for tuned $\lambda_1, \lambda_2$ (solid line) at each $T$; fixed choice $\lambda_1 = 0.89, \lambda_2 = 0.97$ (dashed line) tuned for $T=20$. The numerical inversion  of raw data in the upper right inset introduces artifacts and is unstable for small Avg. SSIM scores. (b)-(e) Columns show single-run maps using a total number of $T=10$ or $T=75$ measurements plotted for the NMQA (b, c) and naive approach (d, e), with a representative control path shown in (b). The numerically tuned parameters are $(\lambda_1, \lambda_2)= (0.89, 0.97)$ and $ (0.93, 0.68)$ for (b),(c) respectively. The single map SSIM values are: (b) 0.37; (c) 0.09; (d) 0.8; (e) 0.47. Other parameters: $d=25$; $T = 5, 10, 15, 20, 25, 50, 75, 100, 125, 250$; $\Sigma_v = 10^{-4}$, $\Sigma_F = 10^{-6}$; true low and high field values of $0.25\pi$ and $0.75\pi$ radians respectively.}    	
\end{figure}

We start with a challenging simulated example, in which $d=25$ qubits are arranged in a 2D grid. The true field is partitioned into square regions with relatively low and high values (\cref{SSIM1_collated}(a), left inset) to provide a clear structure for the mapping procedure to identify. In this case, the discontinuous change of the field values in space means that NMQA will not be able to define a low-error candidate neighborhood for any qubits on the boundary. Both NMQA and naive are executed over a maximum of $T$ iterations such that $t\in[1,T]$.  For simplicity, in most cases, we pick values of $T\geq d$ as multiples of the number of qubits, $d$, such that every qubit is measured the same integer number of times in the naive approach which we use as baseline in our comparison. If $T < d$, the naive approach uniformly randomly samples qubit locations. Both NMQA and the naive approach terminate when $t=T$, and a single-run SSIM score is calculated using the estimated map for each algorithm in comparison with the known underlying ``true'' noise map. The average value of the score over 50 trials is reported as Avg. SSIM. 

The Avg. SSIM score as a function of $T$ is shown in the upper right inset of \cref{SSIM1_collated}(a). For any choice of $T$, \mbox{NMQA} reconstructs the true field with a lower SSIM than the naive approach, indicating better performance. The SSIM score also drops rapidly with $T$ when using NMQA, compared with a more gradual decline in the naive case. Increasing $T\gg d$ leads to an improvement in the performance of both algorithms but also a convergence of the scores as every qubit is likely to be measured multiple times for the largest values considered here.  Representative maps for $T=10$ and $T=75$ measurements under both the naive and NMQA mapping approaches are shown in \cref{SSIM1_collated}(b-e), along with a representative adaptive measurement sequence employed in NMQA in panel (b). In both cases, NMQA provides estimates of the map which are closer to the true field values, whereas naive maps are dominated by estimates at the extreme ends of the range, which is characteristic for simple reconstructions using sparse measurements.  
\FloatBarrier
It is instructive to represent these data in a form that shows the effective performance enhancement of the NMQA mapping procedure relative to the naive approach; the main panel in \cref{SSIM1_collated}(a) reports the improvement in the reduction of  the number of measurements required to reach a desired Avg. SSIM value. The shape of the performance improvement curve is linked to the total amount of information provided to both algorithms. At high Avg. SSIM scores on the far-right of the figure, both naive and NMQA algorithms receive very few measurements and map-reconstructions are dominated by errors. Near the origin, an extremely large number of measurements are required to achieve low Avg. SSIM scores and the ratio of measurements between naive and NMQA tends to unity, corresponding to the convergence mentioned above. In the intermediate regime, a broad peak indicates that NMQA outperforms brute force measurement by up to $18\times$ in reducing total number of qubit measurements over a range of moderate-to-low Avg. SSIM scores for map reconstruction error. Similar performance is achieved for a range of other qubit-array geometries and characteristics of the underlying field (\emph{Supplementary Materials}).

All of our results rely on appropriate tuning of the \mbox{NMQA} particle filter via its parameters $\lambda_1$ and $\lambda_2$. Numerical tuning of these parameters is conducted for each $T$ and this data is represented as solid curves in \cref{SSIM1_collated}(a).  We also demonstrate that using fixed values for these parameters only marginally degrades performance, as indicated by the dashed line in \cref{SSIM1_collated}(a). 

\begin{figure}[t!]
	\includegraphics[scale=1]{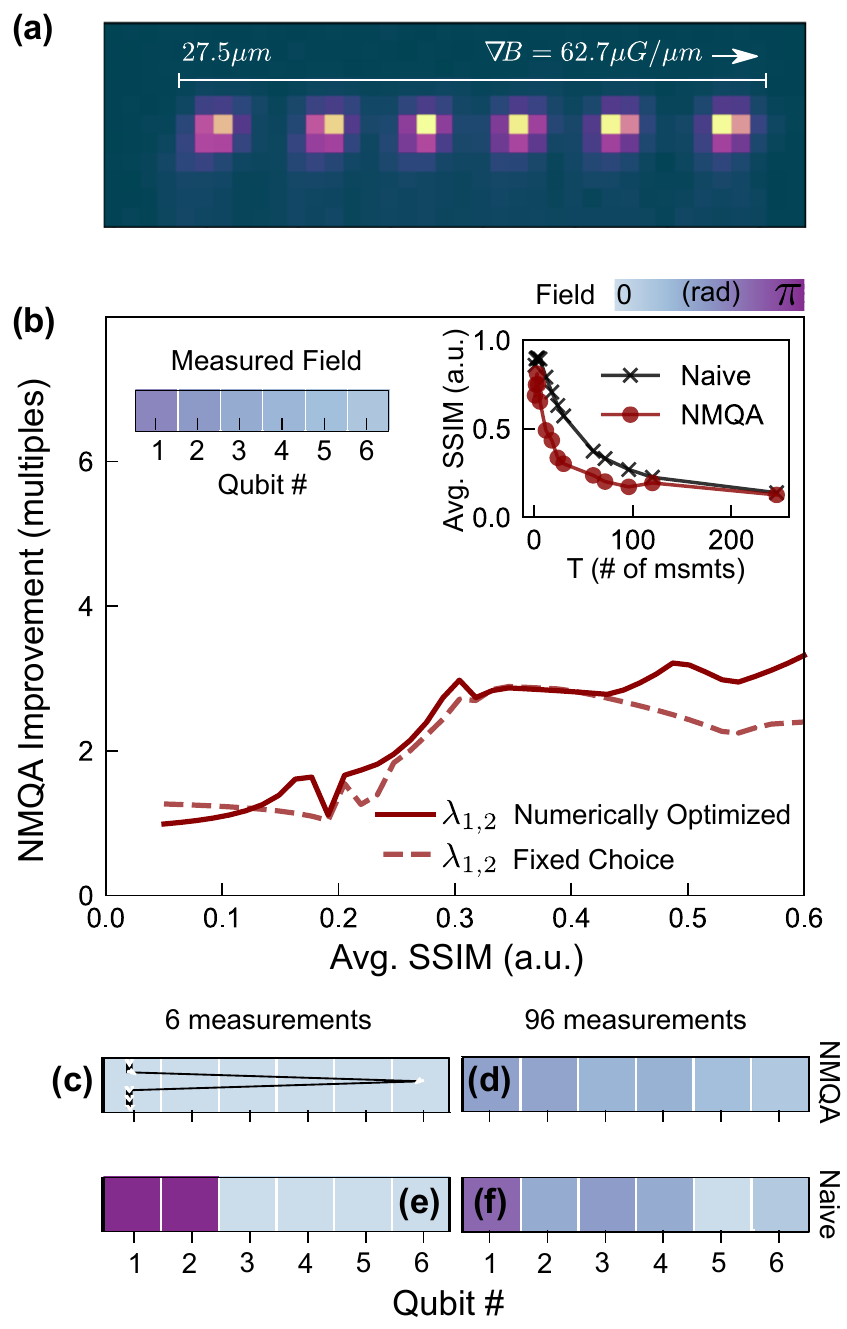}
	\caption{\label{SSIM_expt_0_collated_normal} (a) Image of six trapped ${}^{171}\mathrm{Yb}^{+}$ ions in the fluorescing $|1\rangle$ state averaged over multiple exposures of 750 $\upmu$s. (b) Upper left inset: Color scale indicating a phase shift induced by a magnetic field gradient across 1D array of 6 qubits. Upper right inset: Avg. SSIM score vs total measurement budget $T$ for NMQA and naive algorithms averaged over 50 trials. Main panel: ratio of naive to NMQA measurements vs. Avg. SSIM score for tuned $\lambda_1, \lambda_2$ (solid line) at each $T$; fixed choice $(\lambda_1, \lambda_2) = (0.95, 0.99) $ tuned for $T=24$. The numerical inversion of raw data in the upper right inset introduces artifacts and is unstable for small SSIM scores. (c)-(f) Columns show single-run maps using a total number of $T=6$ or $T=96$ measurements plotted for the NMQA (c, d) and naive (e, f), with a representative control path shown in (c). The numerically tuned parameters are $(\lambda_1, \lambda_2)= (0.97, 0.94)$ for (c) and $(0.97, 0.95)$ for (d). Single map SSIM values are (c) 0.99; (d) 0.35; (e) 0.77; (f) 0.50. Other parameters: $d=6$; $T = 1, 2, 3, 4, 6, 12, 18, 24, 30, 60, 72, 96, 120, 246$; $\Sigma_v = 10^{-4}$, $\Sigma_F = 10^{-6}$; true low and high field values of $0.25\pi$ and $0.75\pi$ radians respectively.}
\end{figure}

We now apply the NMQA mapping algorithm to real experimental measurements on qubit arrays.  In \cref{SSIM_expt_0_collated_normal}, we analyze Ramsey experiments conducted on an array of $^{171}\text{Yb}^{+}$ ions confined in a linear Paul trap, with trap frequencies \mbox{$\omega_{x,y,z}/2\pi \approx (1.6,1.5,0.5)~\text{MHz}$}. Qubits are encoded in the $^2\mathrm{S}_{1/2}$ ground-state manifold where we associate the atomic hyperfine states \mbox{$|F=0, m_F =0\rangle$} and \mbox{$|F=1, m_F =0\rangle$} with the qubit states $\vert0\rangle$ and $|1\rangle$, respectively. State initialization to $|0\rangle$ via optical pumping and state detection are performed using a laser resonant with the $^2\mathrm{S}_{1/2} - {^2\mathrm{P}_{1/2}}$ transition near 369.5~nm. Laser-induced fluorescence (corresponding to projections into state $\vert 1\rangle$) is recorded using a spatially resolving EMCCD camera yielding simultaneous readout of all ions. In this experiment, qubit manipulation is carried out using microwave radiation at 12.6~GHz delivered via an in-vacuum antenna to all qubits globally.

The trapped ions experience an uncontrolled (``intrinsic''), linear magnetic field gradient which produces spatially inhomogeneous qubit frequencies over the trap volume. When manipulated using a global microwave control field, this results in a differential phase accumulation between qubits. The magnitude of the gradient is illustrated in \cref{SSIM_expt_0_collated_normal}(a), superimposed on an image of six ions in the bright state $\vert 1\rangle$. We aim to probe this field gradient through the resulting qubit detuning and associated differential phase accumulation throughout each measurement repetition.

A preliminary Ramsey experiment, in which the interrogation time is varied and fringes observed, confirms that at an interrogation time of $40$ms accumulated relative phase in a Ramsey measurement is $<\pi$ radians. A total of $25,500$ Ramsey measurements with a wait time of $40$ ms are performed on all six ions in parallel.  For each repetition, a standard  machine-learning image classification algorithm assigns a `0' (dark) or `1' (bright) to each ion based on a previously recorded set of training data. From averaging over repetitions, we construct a map of the accumulated phase (and hence the local magnetic field inducing this phase) on each qubit, shown schematically in the right inset of \cref{SSIM_expt_0_collated_normal}(b). We consider the field extracted from this standard averaging procedure over all repetitions of the Ramsey experiment, at each ion location, as the ``true'' field against which mapping estimates are compared using the SSIM score as introduced above.

We employ the full set of $6 \times 25,500$ measurements as a data-bank on which to evaluate and benchmark both algorithms.  At each iteration, the algorithm determines a measurement location in the array and then randomly draws a single, discretized measurement outcome (0 or 1) for that ion from the data-bank.  The rest of the algorithmic implementation proceeds as before.  Accordingly, we expect that the naive approach must smoothly approach a SSIM score of zero as $T$ increases (\cref{SSIM_expt_0_collated_normal}(b), left inset). 
In these experiments, we experience a large measurement error arising from the necessary detection time of 750~$\upmu$s associated with the relatively low quantum efficiency and effective operating speed of the EMCCD camera. Performing spatially resolved measurement across the qubit array introduces a large measurement error due to the low quantum efficiency of the EMCCD camera.  Compensating this via extension of the measurement period leads to an asymmetric bias due to state decays that occur during the measurement process in $\approx1.5$~ms ($\vert 1\rangle \rightarrow \vert 0 \rangle$) and  $\approx30$~ms ($\vert 0\rangle \rightarrow \vert 1 \rangle$) under the laser power and quantization magnetic field strength used in our experiment. In our implementation, neither NMQA nor the brute-force algorithm was modified to account for this asymmetric bias in the detection procedure, although in principle the image classification algorithm employed to determine qubit states from fluorescence detection can be expanded to account for this.  Therefore, both NMQA and the naive approach are affected by the same detection errors. Despite this complication, we again find that NMQA outperforms the naive mapping algorithm by a factor of $2-3$ in the number of required measurements (\cref{SSIM_expt_0_collated_normal}(b)), with expected behavior at the extremal values of Avg. SSIM.

\section{Conclusion \label{sec:conclusion}}

In this work we presented NMQA - a framework for autonomous learning where we reconstruct an unknown spatial noise field by efficiently scheduling measurements on multi-qubit devices.  We developed a novel, iterative, maximum-likelihood procedure implemented via a two-layer particle filter to share state-estimation information between qubits within small spatial neighborhoods, via a mapping of the underlying spatial correlations of the noise field. An autonomous controller schedules future measurements in order to reduce the estimated uncertainty of the map reconstruction.  Numerical simulations and calculations run on real experimental data demonstrated that NMQA outperforms a naive mapping procedure by reducing the required number of measurements up to $3\times$ in achieving a target map similarity score on a 1D array of trapped ytterbium ions (up to $18\times$ in 2D using simulated data).

Beyond these example demonstrations, the key numerical evidence for the correctness of NMQA's functionality came from the observation that the tuned values of $\lambda_1, \lambda_2 \gg 0 $ for all results reported here. Since numerically tuned values for these parameters were found to be non-zero, we conclude information sharing in NMQA is non-trivial and the algorithm departs substantially from a  brute-force measurement strategy. This contributes to the demonstrated improvements in Avg. SSIM scores when using NMQA in \cref{SSIM1_collated,SSIM_expt_0_collated_normal}. 

Overall, achieving suitable autonomous learning through NMQA requires that the $(\lambda_1, \lambda_2)$ parameters are appropriately tuned using measurement data-sets as part of a training procedure. To select $(\lambda_1, \lambda_2)$, we chose a $(\lambda_1, \lambda_2)$ pair with the lowest expected error. Such an error calculation requires apriori information about the measured (or true) field, with the expected value computed over many runs. Numerical analysis shows that tuned $\lambda_1, \lambda_2 \in [0.5, 1]$ (top right quadrant of the unit square) and this observation holds for a range of different noise fields with regularly spaced 1D or 2D configurations. These numerical observations suggest that theoretical approaches to deducing `optimal' regions for $(\lambda_1, \lambda_2)$ may be possible.  Meanwhile, in a specific application, a practitioner may only have access to the real-time rate of change of state estimates and residuals, {\em i.e.} a particle filter generates a predicted measurement at location $j$ which can be compared to next the actual measurement received at $j$. Monitoring the real-time rate of change of state estimates and/or residuals in a single run can be used to develop effective stopping criteria to set $(\lambda_1, \lambda_2)$ with minimal apriori knowledge of the physical system. These extensions to the operability of NMQA are active research areas.

The framework we have introduced is flexible and can accommodate temporal variations in the system such as drifts in the map and changes in the availability of sensor qubits.  We are also excited to explore how exploitation of hardware platforms in which qubit locations are not rigidly fixed during fabrication, such as with trapped-ions, may allow sub-lattice spatial resolution by dynamically translating individual qubits during the mapping procedure. Our work is part of an exciting new area of future study exploring the intersection between hardware architectures and control solutions~\cite{BallPRApplied2016} in NISQ-era quantum computers, and may also have bearing on distributed networks of quantum sensors.

\section*{Methods}

We summarize the structure of the NMQA algorithm in \cref{algorithm:q-slam-pf} using a pseudocode representation. 

The first part of the algorithm consists of an initialization procedure which ensures all particles are sampled from the prior for extended state vector at $t=0$, giving $X_0$. All particles are equally weighted before measurement data is received. 
\begin{algorithm}[H] 
	\caption{NMQA}\label{algorithm:q-slam-pf}
	\begin{algorithmic}[0] 
		\Procedure{NMQA}{$d$ qubit locations, $\lambda_1, \lambda_2$}
		\\
		\If{$ t = 0$}
		\Procedure{Initialize}{$X_0$} 
		\For{$ \alpha \in \{1, 2, \hdots, n_\alpha \}$}
		\State Initially sample $x_{0}^{(\alpha)} \sim \pi_0$ 
		\State Initially compute $W_0^{(\alpha)} = \frac{1}{n_\alpha}$
		\EndFor
		\EndProcedure 
		\EndIf
		\\
		\While{$ 1 \leq t < T$}
		\If{Controller}
		\State $j_t, Y_t^{(j_t)} \gets$ \Call{Controller}{$X_{t-1}$} \Comment{Qubit $j$, at $t$}
		\EndIf
		\For{$ \alpha \in \{1, 2, \hdots, n_\alpha \}$}
		\State $\{x_{t}^{(\alpha)}\} \gets $ \Call{PropagateStates}{$\{x_{t-1}^{(\alpha)} \}$ }
		\State Update $F_t^{(\cdot), (\alpha)}$ via $ \{Y_t^{(j_t)},  \{\hat{Y}_t^{(q_t)}\}, \lambda_1\} $
		\State $\{\{x_t, W_t\}^{(\alpha, \beta_\alpha)}\} \gets $ \Call{ComputeWeights}{$\{x_{t}^{(\alpha)}\}$}
		\State $\{x_t^{(\alpha, \beta_\alpha)}, \frac{1}{n_\alpha n_\beta}\} \gets $ \Call{Resample}{$\{\{x_t, W_t\}^{(\alpha, \beta_\alpha)}\}$}
		\State Update $R_t^{(j_t), (\alpha)}$
		\State $\{\{x_t, W_t\}^{(\alpha)}\} \gets$ \Call{Collapse$\beta$}{$\{x_t^{(\alpha, \beta_\alpha)}, \frac{1}{n_\alpha n_\beta}\}$}
		\State $\{x_t^{(\alpha)}, \frac{1}{n_\alpha}\} \gets $ \Call{Resample}{$\{\{x_t, W_t\}^{(\alpha)}\}$}
		\EndFor
		\State $\{\hat{Y}_{t+1}^{(q)}\}_{q\in Q_{t+1}} \gets $ \Call{Generate$\hat{Y}$}{Posterior $X_t$}
		\EndWhile \label{pseudoalgo:NMQAr:endwhile2}			
		\EndProcedure
		\\ \dotfill
		\Function{ComputeWeights}{$\{x_{t}^{\alpha}\}$}
		\For{$ \alpha \in \{1, 2, \hdots, n_\alpha \}$}
		\State Compute $\tilde{W}_t^{( \alpha )} = g_1(\lambda_1, Y_t^{(j)}) $ 
		\State $\{x_t^{(\alpha, \beta_\alpha)}\} \gets $  Generate $\beta$-layer
		\For{$\beta_\alpha \in \{1, 2, \hdots, n_\beta \} $}
		\State Compute $\tilde{W}_t^{( \beta_\alpha | \alpha)} = g_2(\lambda_2, Q_t) $
		\EndFor
		\State Normalize $\tilde{W}_t^{( \beta_\alpha | \alpha)}$
		\EndFor
		\State Normalize $\tilde{W}_t^{( \alpha)}$
		\State Compute $W_t^{(\alpha, \beta_\alpha)} =  \tilde{W}_t^{( \beta_\alpha | \alpha)} \tilde{W}_t^{( \alpha)} \quad \forall \{\alpha, \{\beta_\alpha\} \}$
		\State Return $ n_\alpha n_\beta $ particles and weights $\{\{x_t, W_t\}^{(\alpha, \beta_\alpha)}\}$		
		\EndFunction
	\end{algorithmic}	
\end{algorithm}
For $t>0$, the function \texttt{PropagateStates} represents the transition probability distribution for Markov $X_t$ i.e. it represents identity dynamics and is a placeholder for future extensions to model dynamical $F_t$. In each $t$, a single physical measurement is received. This triggers a set of update rules for $F_t$ and $R_t$. We note that the state variables, $F_t$ and $R_t$, are updated in a specific order within each time-step $t$. The order of these computations correspond to the iterative maximum likelihood approximation for the NMQA framework. The form of the update rules reflect NMQA's unique state vector and NMQA's information sharing procedures. Map and neighborhood information is carried via different particle types, and information sharing is implemented via particle updates and re-sampling steps. For each type of particle, the weights are computed according to NMQA likelihood functions $g_1(\lambda_1, Y_t^{(j)})$ and $g_2(\lambda_2, Q_t)$.

Despite atypical computational steps and unique likelihood functions, the total number of particles is constant at the beginning and end of each time-step $t$ and the particle branching mechanism for NMQA remains a multi-nomial branching process. These multi-nomial branching processes are characteristic of standard particle filtering techniques \cite{bain2009}. 

\subsection{Structural Similarity Metric Definition}

For all analysis, we use a risk metric called the Structural Similarity Index (SSIM) \cite{wang2004image}. This metric is used to conduct optimization of NMQA parameters and assess performance relative to the naive measurement strategy. For two vectorized images $x$ and $y$, the metric is defined as:
\begin{align}
	s(x, y) & = \frac{(2 \mu_x\mu_y +C_1)(2\sigma_{xy} + C_2)}{(\mu_x^2  + \mu_y^2 + C_1 )(\sigma_x^2 + \sigma_y^2 + C_2)} \\
	\text{SSIM}(x, y) &:= |1 - s(x,y)|
\end{align} In the formula above, $\mu_i, \sigma_i^2, i = x,y$ represent the sample estimates of the means and variances of the respective vectorized images, and $\sigma_{xy}$ captures correlation between images. The term $s(x,y)$ is the key metric developed in \cite{wang2004image} and it includes arbitrary constants $C_1=C_2=0.01$ which stabilize the metric for images with means or variances close to zero. The ideal score given by $s(x, y)$ is unity, and corresponds uniquely to the case $x=y$. We report the absolute value of the deviations from the ideal score of unity, where the direction of the deviation is ignored as given by $\text{SSIM}(x, y)$. For our application, this $\text{SSIM}(x, y)$ metric lies between $[0, 1]$ (negative values of $s(x, y)$ are not seen in our numerical demonstrations). We report the average of $\text{SSIM}(x,y)$ values over 50 trials as $\text{Avg. SSIM}$.

\section*{Acknowledgments}
Authors thank V.M. Frey for proposing methods for single-ion state detection using camera images, and S. Sukkarieh, A.C. Doherty, and M. Hush for useful discussions. This work partially supported by the ARC Centre of Excellence for Engineered Quantum Systems CE170100009, the US Army Research Office under Contract W911NF-12-R-0012, and a private grant from H. \& A. Harley.\\

\section*{Data Availability}
All simulated and experimental data to reproduce all figures can be accessed via links in the Supplementary Materials without restrictions.\\

\section*{Code Availability}
Minimally reproducing equations for NMQA and links to the code-base are provided in the Supplementary Materials without restrictions.\\

\section*{Author Contributions}
The NMQA theoretical framework and numerical implementations were devised by R. Gupta based on research directions set by M.J. Biercuk. R. Gupta and M.J. Biercuk co-wrote the paper. A. Milne, C. Edmunds and C. Hempel led all experimental efforts and contributed to the paper draft.\\

\section*{Competing Interests}
The authors declare that there are no competing interests.

\section{Materials and Correspondence} 
Correspondence to R. Gupta.



%

\end{document}


\title{Supplementary Materials}
\maketitle

The Supplementary Materials provide additional performance results to support the main findings of the paper. We  also provide the minimal set of equations required to implement NMQA  as a non-linear, particle filter. These equations specify the computational steps to accompany the \textit{Methods} section of the main text.

\section{Supporting Results}
In this section, we provide details about the optimization approach used to produce all data-figures. We also report additional simulation results for various underlying spatial fields and qubit configurations.

\subsection{$(\lambda_{1}, \lambda_{2})$ Optimization}

The numerical optimization of NMQA parameters, $(\lambda_1, \lambda_2)$, helps assess whether the NMQA algorithm provides meaningful adaptive measurement scheduling, relative to a brute force measurement strategy, for any specific practical application.

\begin{figure}[H]
	\includegraphics[scale=1]{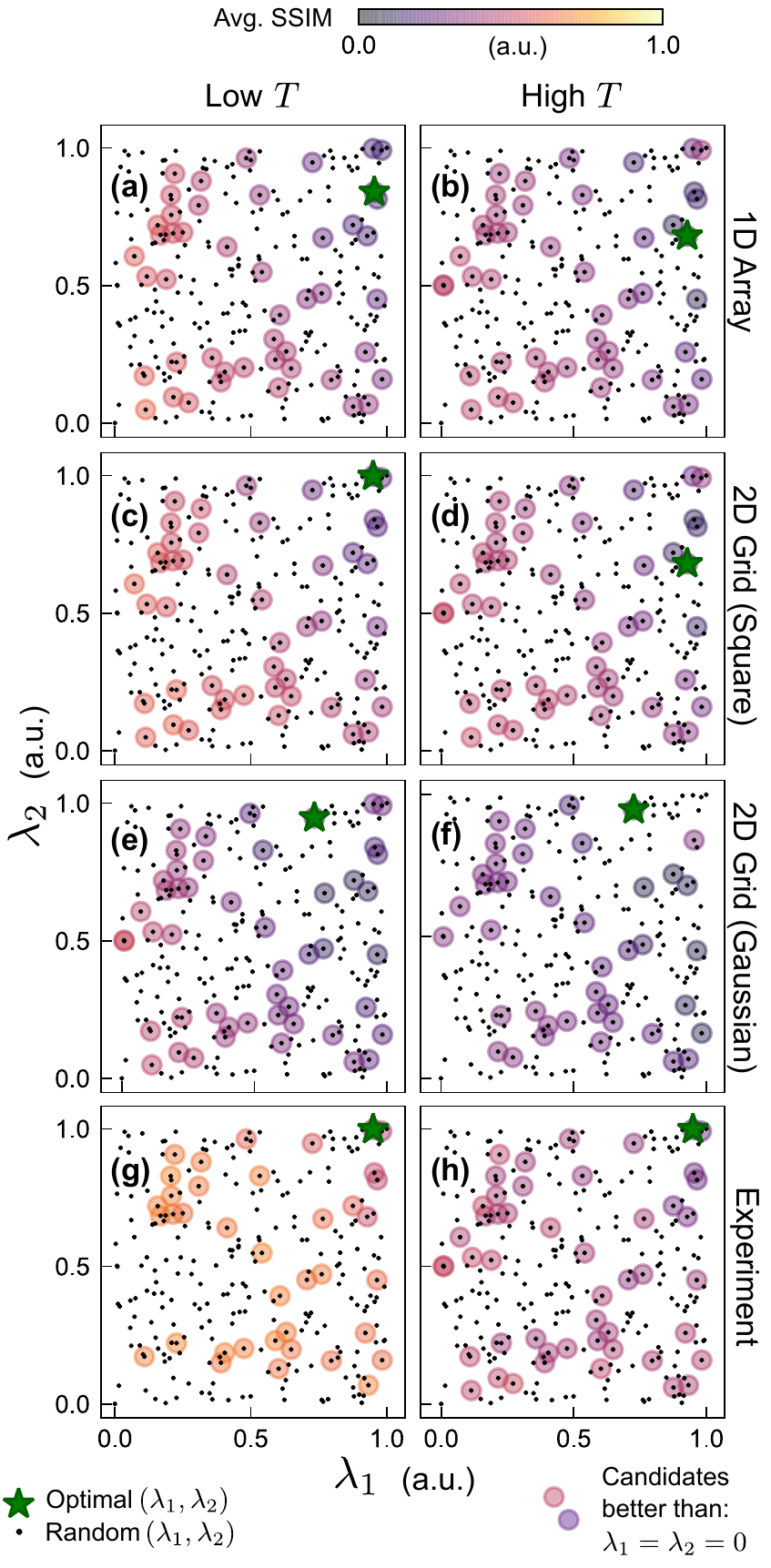}
	\caption{\label{SSIM_optimisation} Numerical optimization of $(\lambda_1, \lambda_2)$. Rows correspond to different physical configurations; columns correspond low $T$ regime [left] and high $T >> d$ regime [right]. (a)-(f) correspond to NMQA analysis with simulated data, $d=25$; (g),(h) correspond to experimental measurements, $d=6$. Each panel depicts 250 randomly chosen $(\lambda_1, \lambda_2)$ pairs (black dots). Shaded circles highlight $(\lambda_1, \lambda_2)$ pairs with lower Avg. SSIM score compared to $\lambda_1= \lambda_2 =0$ case; color-scale conveys actual Avg. SSIM value. Optimal $(\lambda_1, \lambda_2)$ pair has lowest Avg. SSIM score (green star): (a) $(0.80, 0.96)$, $T=25$; (b) $(0.93, 0.68)$, $T=125$; (c) $(0.89, 0.97)$, $T=20$; (d) $(0.84, 0.96)$, $T=125$; (e) $(0.72, 0.84)$, $T=25$; (f) $(0.77, 0.97)$, $T=125$; (g) $(0.95, 0.99)$, $T=24$; (h) $(0.97, 0.95)$, $T=96$. Other parameters: $\Sigma_v = 10^{-4}$, $\Sigma_F = 10^{-6}$, $50$ trials.}
\end{figure}

In \cref{SSIM_optimisation}, we plot the $\text{Avg. SSIM}$ score for 250 randomly sampled $(\lambda_1, \lambda_2)$ pairs and we compare this with the expected value of the $\text{Avg. SSIM}$ score for $\lambda_1, \lambda_2 = 0$ in \cref{SSIM_optimisation}. The first three rows of \cref{SSIM_optimisation} correspond to three different simulations of the true field: a 1D step field of \cref{SSIM0_collated}, a 2D square field reported in the main text as Fig.~3, and a 2D Gaussian field corresponding to \cref{SSIM2_collated}. The last row corresponds to results using experimental measurements presented in Fig.~4 of the main text. 

In each panel, a black dot represents a randomly selected $(\lambda_1, \lambda_2)$ pair. Shaded circles select a pair with an Avg. SSIM score better than $\lambda_1, \lambda_2 = 0$ by at least 0.025 in arbitrary units (2.5\% of maximal expected deviation of unity). The color of the shaded circle refers to the actual value of the $\text{Avg. SSIM}$, given by the color bar at the top of \cref{SSIM_optimisation}, with darkest purple values representing the ideal score of zero. The green star indicates the lowest score achieved over all pairs, and defines the numerically tuned $(\lambda_1, \lambda_2)$ values used for analysis. This procedure is repeated for all $T$ in the solid lines of \cref{SSIM0_collated,SSIM2_collated} and Figs.~3 and 4 of main text.

The fixed choice case refers to finding a numerically optimal $(\lambda_1, \lambda_2)$ pair for a fixed (low) $T$ value, and using this pair for all other choices of $T$, including the high $T$ regime where $T \gg d$. The fixed choice $(\lambda_1, \lambda_2)$ pair is given by green stars in the leftmost column of \cref{SSIM_optimisation}. For simulations, $T$ is fixed to $T \approx d$. For experiment in Fig.~4 of main text, $T$ is set as $T=3d$ to accommodate increased measurement and detection error. High $T$ regimes correspond to $T \gg d$, with $T$ up to $5-15\times$ greater than system size, $d$. 

We apply the tuned $(\lambda_1, \lambda_2)$ pairs (green stars) in the first column of \cref{SSIM_optimisation} to all choices of $T$, including $T \gg d$, yielding the dashed crimson lines in \cref{SSIM0_collated,SSIM2_collated} and Figs.~3 and 4 of main text. To aid visual comparisons, we use the same set of random pairs $(\lambda_1, \lambda_2)$ across all experiments, but results hold if candidate parameters are randomized across experiments. For the numerical demonstrations in the main text and the Supplement, it is seen that numerically tuned $\lambda_1, \lambda_2 \neq 0$ and NMQA sharing is non-trivially contributing to the overall inference procedure.

Of the experiments reported in \cref{SSIM_optimisation}, one observes that performance improves from bottom left to top right corners of all panels, indicating a performance improvement away from zero values for both parameters. With $\lambda_1, \lambda_2 \equiv 1$, shared information in NMQA is treated on an equal footing with information obtained from physical data. For $\lambda_1, \lambda_2 \equiv 0$, NMQA effectively reduces to a brute force measurement strategy but with uniformly randomly sampled locations. Since performance appears to improve as the two parameters, $\lambda_1, \lambda_2$ move away from zero and this trend is unchanged even as the total amount of measurement data increases (left column to right), we interpret this as numerical evidence that the sharing mechanism in NMQA framework is both non-trivial and correct in the large data limit.

\subsection{Additional Numerical Simulations}
The performance of NMQA is of course conditioned on the spatial variation in the true field relative to the configuration of the qubit-grid. 

\begin{figure}[t!]
	\includegraphics[scale=1]{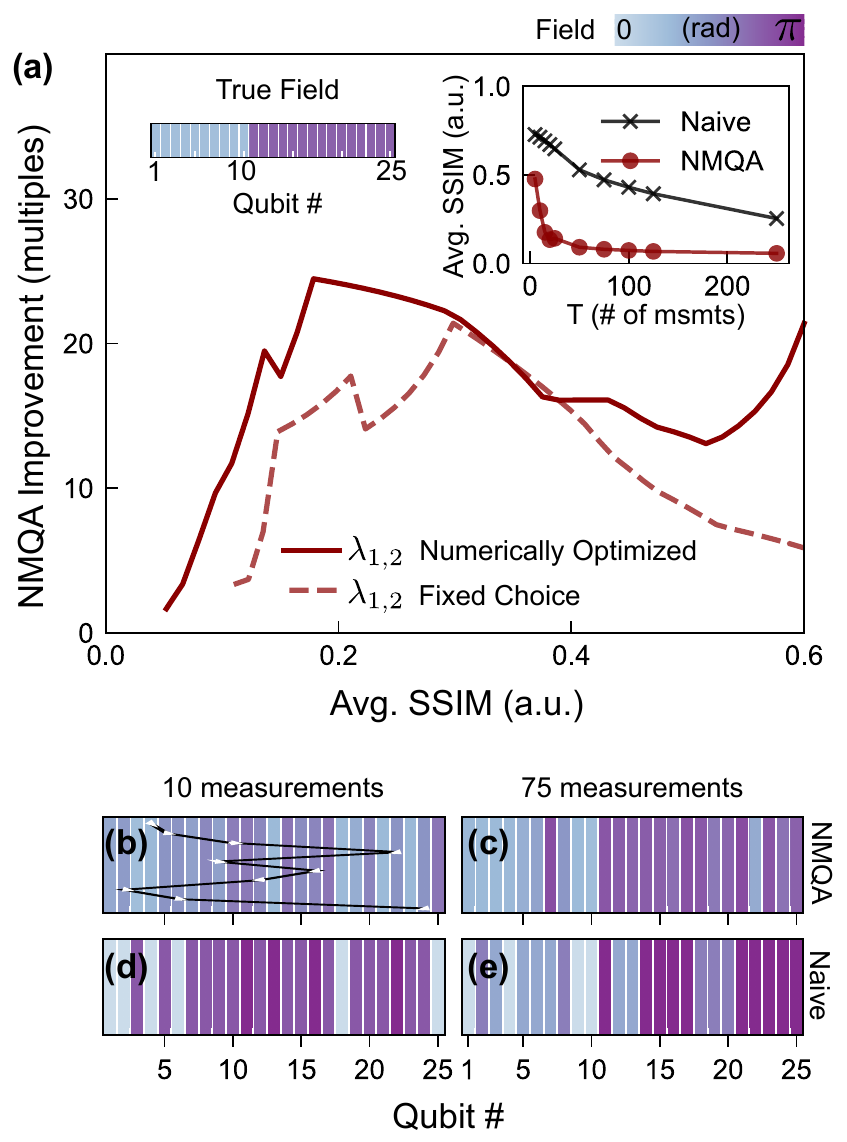}
	\caption{\label{SSIM0_collated} (a) Upper left inset:  1D array of 25 qubits in `step' field shown as a colorscale. Upper right inset: Avg. SSIM score vs total measurement budget $T$ for NMQA and naive algorithms averaged over 50 trials. Main panel: ratio of naive to NMQA measurements vs. Avg. SSIM score for tuned $\lambda_1, \lambda_2$ (solid line) at each $T$; fixed choice $\lambda_1 = 0.80, \lambda_2 = 0.96$ (dashed line) tuned for $T=25$. The numerical inversion of raw data in the upper right inset introduces artifacts and is unstable for small SSIM scores. (b)-(e) Columns show single-run maps using a total number of $T=10$ or $T=75$ measurements plotted for the NMQA (b, c) and naive approach (d, e), with a representative control path shown in (b). The numerically tuned parameters are  $(\lambda_1, \lambda_2)= (0.80, 0.96)$ and $ (0.93, 0.68)$ for (b),(c) respectively. The single map SSIM values are: (b) 0.64; (c) 0.12; (d) 0.76;  (e) 0.49. Other parameters:  $d=25$; $T = 5, 10, 15, 20, 25, 50, 75, 100, 125, 250$; $\Sigma_v = 10^{-4}$, $\Sigma_F = 10^{-6}$; true low and high field values of $0.25\pi$ and $0.75\pi$ radians respectively.}    
\end{figure}

\begin{figure}[H]
	\includegraphics[scale=1]{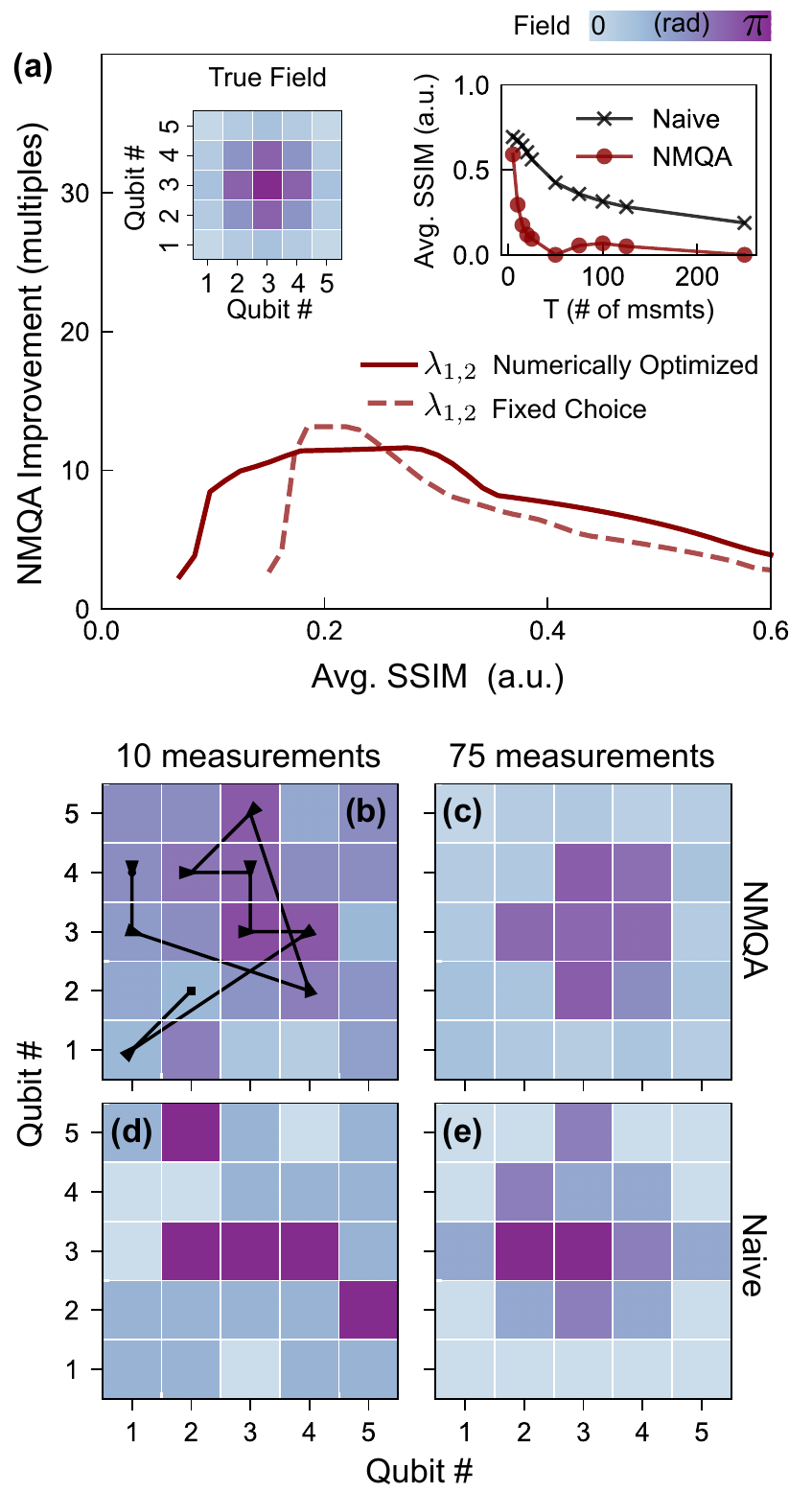}
	\caption{\label{SSIM2_collated} (a) Upper left inset: 2D array of 25 qubits in $5 \times 5$  `Gaussian' field shown as a colorscale. Upper right inset:  Avg. SSIM score vs total measurement budget $T$ for NMQA and naive algorithms averaged over 50 trials. Main panel: ratio of naive to NMQA measurements vs. Avg. SSIM score for tuned $\lambda_1, \lambda_2$ (solid line) at each $T$; fixed choice $\lambda_1 = 0.72, \lambda_2 = 0.84$ (dashed line) tuned for $T=25$. The numerical inversion of raw data in the upper right inset introduces artifacts and is unstable for small SSIM scores. (b)-(e) Columns show single-run maps using a total number of $T=10$ or $T=75$ measurements plotted for the NMQA (b, c) and naive approach (d, e), with a representative control path shown in (b). The numerically tuned parameters are $(\lambda_1, \lambda_2)= (0.78, 0.81)$ and $ (0.81, 0.63)$ for (b),(c) respectively. The single map SSIM values are: (b) 0.16; (c) 0.01; (d) 0.67;  (e) 0.27. Other parameters: $d=25$; $T = 5, 10, 15, 20, 25, 50, 75, 100, 125, 250$; $\Sigma_v = 10^{-4}$, $\Sigma_F = 10^{-6}$; true low and high field values of $0.25\pi$ and $0.75\pi$ radians respectively.}
\end{figure}

\cref{SSIM0_collated} provides a simulation of the performance of NMQA when neighbours are constrained to a line in 1D. This 1D case can be compared with Fig.~3 of the main text, where we re-stack 25 qubits from 2D to 1D while keeping the overall ratio of qubits subject to low and high field regions the same. By reducing spatial dimensions on the qubit-grid, we influence neighbourhood selection within the algorithm. 

In \cref{SSIM2_collated}, we provide an example of a true field where the value of the field changes by almost every `pixel' that is, spatial variation of the true field is `fast' compared to results in Fig.~3 of main text. From standard classical sampling considerations, we expect that the spatial variation in the true field relative to the inter-qubit spacing in any dimension will affect overall performance.

In these additional simulations, NMQA still outperforms naive by $2\times$-$15\times$ for a large range of error scores and the ratio between NMQA and naive measurements approaches unity in the large data limit.
\FloatBarrier
\section{Minimally Reproducing Equations for NMQA Filtering Problem}

This section introduces mathematical details for the NMQA non-linear filter. We provide definitions, equations and the order of computations required to reproduce the algorithm. 

A detailed introduction to these mathematical objects and computations; supporting derivations, and an analysis of their properties using standard non-linear filtering theory, will be provided in a forthcoming technical manuscript.
\subsection{Physical Measurement Model}
Our system consists of $d$ qubits and a total number of $T$ measurements are taken at $t=1, 2, \hdots, T$. Qubit locations are labeled by $\{1, 2, \hdots d\}$. A measurement $Y_t^{(j_t)} \in \{0,1\}$ is a random variable obtained by measuring a single qubit at the location $j_t \in \{1, 2, \hdots d\}$ at iteration $t$. 

The set of phases for all qubits in the system is the random vector $F_t$ and it is referred to as the `true map' or equivalently, as the set of `map values'. This state vector $F_t \in \mathbb{S}_F := [0, \pi]^d$ takes in values between $[0, \pi]$ and the map-value, $F_t^{(j_t)}$ is a noise-induced Born phase for a qubit at location $j_t$. Here, and elsewhere, the notation $\mathbb{S}$ refers to a complete, separable metric space and $\mathcal{S} \equiv \ofieldgen{\mathbb{S}}$ is the associated Borel $\sigma$-algebra generated by $\mathbb{S}$ for a particular random variable, in this case, for $F$. These definitions are required to specify spaces of probability measures and continuous, bounded, Borel-measurable functions for particle filtering methods.

The measurement model for our state vector incorporates a projective single-qubit measurement. For a given location $j_t$ and iteration $t$, the measurement model is:
	\begin{align}
	Y_t^{(j_t)} := \mathcal{Q}(\frac{1}{2} \cos(F_t^{(j_t)}) + v_t + \frac{1}{2}) \label{qslam:measurementmodel}
	\end{align}
Here, $\mathcal{Q}$ represents a Bernoulli trial parameterized by a probability that depends on the map-value $F_t^{(j_t)}$ and a noise term, $v_t$. 

This measurement noise term has a form resembling a truncated Gaussian error model with zero mean and $\Sigma_v$ variance; but it is derived from a study of amplitude-quantised sensor error in classical signal processing. The measurement model leads to a likelihood function given by:
	\begin{align}
	 g_1(\lambda_1, Y_t^{(j_t)}) & :=  \frac{\rho_0}{2} \nonumber \\
	& + \frac{\rho_0  \cos(F_t^{(j_t)}) }{2} \left( \delta(Y_t^{(j_t)} - 1) - \delta(Y_t^{(j_t)}) \right) \label{qslam:likelihood:g1:func} \\
	\rho_0 &: =  \erf(\frac{2b}{\sqrt{2\Sigma_v}}) + \frac{\sqrt{2\Sigma_v}}{2b} \frac{e^{-(\frac{2b}{\sqrt{2\Sigma_v}})^2}}{\sqrt{\pi}}  - \frac{1}{2b}\frac{\sqrt{2\Sigma_v}}{\sqrt{\pi}}\label{qslam:likelihood:g1:rho0}  \\
	b & := 1/2 
	\end{align} In the likelihood function above, the parameters $\rho_0, b$ are set by the properties of measurement noise when sensor data is allowed to take on only discrete binary values. In practice, the term $F_t^{(j_t)}$ depends on $\lambda_1$ under certain particle-filtering approximations, as introduced in the final section. The prior for $F_0$ is uniformly distributed.

\subsection{Information Sharing Mechanism}
Next, we introduce the mechanism to share state information between qubits within a local neighbourhood. We use the concept of `blurring' state information from a point value at $j_t$ into a small neighbourhood  around $j_t$, and the set of neighbouring qubits in this region is given by $Q_t$. Both the size of the neighbourhood and the interpolation of map values within the neighborhood is parameterized by  a random length-scale variable, $R_t^{(j_t)}$, at $j_t$. 
For some $R_{min} \in \mathbb{R} > 0$, we define the state vector $R_t \in  \mathbb{S}_R := [R_{min}, R_{max}]^d$ the set of random variables associated with $d$ qubit locations, taking in values between $[R_{min}, R_{max}]$, such that $R_t^{(j_t)}$ is the $j_t$ element of $R_t$. The length-scale parameterizes a Gaussian function with amplitude $F_t^{(j_t)}$  that spreads the point-value of the phase $F_t^{(j_t)}$, at $j_t$, into a small neighbourhood. Let $\nu_{(j_t,q)}$ be the separation distance between the qubit at $j_t$ and a point $q$ in a neighbourhood around $j_t$. Then the smeared phase value is given by:
	\begin{align}
	\bar{F}(\nu_{(j_t,q)}) &: = F_t^{(j_t)} \exp\left( \frac{- \nu_{(j_t,q)}^2}{(R_t^{(j_t)})^2}\right) \label{qslam:defn:bar_F}
	\end{align}
The point value of the phase  is regained in the limit $R_t^{(j_t)} \to 0$, and the smallest spatial resolution sets the value of $R_{min}$.  

The neighbourhood, $Q_t$ of the measured qubit at $j_t$ with length-scale $R_t^{(j_t)}$ is the set:
\begin{align}
Q_t &:= \{ q_t | \nu_{(j_t, q_t)} \leq k_0 R_t^{(j_t)}\nonumber \\
& \forall q_t \in \{1, 2, \hdots d \} \setminus \{j_t\} \}, \quad k_0 \in [1, \infty). \label{qslam:defn:Q_t} 
\end{align} That is, neighbours are the set of qubits whose separations distances fall within a radius of $k_0 R_t^{(j_t)}$ about the location $j_t$. Here, $k_0$ is an arbitrary constant that truncates the neighbourhood when smeared phase values at the boundary of the neighborhood are dissimilar to those at the center. A conservative approach is to set $k_0=1$. The index $q_t$ denotes the location labels for all qubits excluding the measured qubit, $ q_t \in \{1, 2, \hdots d \} \setminus \{j_t\} $.

Given a map $F_t$, we associate a likelihood function for the length-scale at $j_t$, $R_t^{(j_t)}$, that scores the most appropriate length-scale for sharing information between qubits. For a constant $\lambda_2 \in [0,1]$, and a non-negative natural number $\tau_{q_t}$, the likelihood for the length-scale at $j_t$ compares the estimate $\mathcal{X}_{q_t} $ to the best available phase information on each neighbour:
	\begin{align}
		g_2(\lambda_2, Q_t) & := \prod_{q_t \in Q_t} \frac{1}{k_1\sqrt{2\pi \Sigma_F}} \exp \left( -\frac{( F_t^{(q_t)} - \mathcal{X}_{q_t} - \mu_F )^2}{2 \Sigma_F }\right) \label{qslam:defn:likelihood:Rt} \\
		F_t^{(q_t)} &= \mathcal{X}_{q_t} + w_t, \quad \forall q_t \in Q_t  \label{qslam:defn:msmtmodel:Rt}\\
		\mathcal{X}_{q_t} &:= (1 - \lambda_2^{\tau_{q_t}})F_t^{(q_t)} + \lambda_2^{\tau_{q_t}} \bar{F}(\nu_{(j_t,q_t)}) \label{qslam:defn:msmtmodel:Rt:Xqt} \\
		w_t & \sim \mathcal{N}(\mu_F, \Sigma_F) \\
		k_1&:= \frac{1}{2}\left(\erf(\frac{\pi + \mu_F}{\sqrt{2\Sigma_F}}) + \erf(\frac{\pi - \mu_F}{\sqrt{2\Sigma_F}})\right)
	\end{align} The quantity $\mathcal{X}_{q_t}$ is based on a weighted average of both current map estimate on the neighbouring qubit and state estimates of physically measured qubit at $j_t$, mediated by a factor $\lambda_2^{\tau_{q_t}}$. The noise parameters $ \mu_F, \Sigma_F$ represent the true error in approximating a continuously varying spatial field with overlapping Gaussian functions. This error is Gaussian distributed and error values lie in the finite interval $[-\pi, \pi]$, giving rise to the constant $k_1$. The non-negative $\lambda_2$ is a parameter for the filtering problem, and the constant $ \tau_{q_t} \leq T$  is the tally of the total number of times $q_t$ has been physically measured, as defined in the final section. The prior for $R_t$  is uniformly distributed.

With the definitions above, posterior state information for map-values and length-scales can be defined under a particle-filtering framework. We now generate binary data messages on the basis of posterior state information at $t$, which influences the inference procedure at $t+1$. For a posterior state estimate $F_t$ and $R_t^{(j)}$ at iteration $t$, a physical measurement at $j_t$ is said to \textit{induce} information exchange in the posterior neighbourhood $Q_t$ (equivalently, prior at the start of $t+1$ step before any information updates commence). In notation, these data messages are the random variables $\hat{Y}_{t+1}^{(q_{t+1})} \in \{0,1\}$ generated using the model:
 	\begin{align}
 	\hat{Y}_{t+1}^{(q_{t+1})}  & := \mathcal{Q}(\frac{1}{2} \cos(\mathcal{X}_{q_t}) + \frac{1}{2}) \label{qslam:defn:quasimsmts:1}
 	\end{align} The shared information uses a weighted average of both posterior map estimate on the neighbouring qubit and posterior state information for the physically measured qubit at $j_t$, mediated by a factor $\lambda_2^{\tau_{q_t}}$. Note that $\mathcal{X}_{q_t}$ is to be interpreted as the posterior \cref{qslam:defn:msmtmodel:Rt:Xqt}, where $ q_{t+1} \in Q_{t+1}$ (above) equals the $Q_t$ set by posterior state estimate $R_t^{(j)}$ and the posterior $F_t$ is used for all calculations. 

\subsection{Nonlinear Filtering}
We are now in a position to state the non-linear filtering problem for NMQA in terms of a state variable $X$ and an observation vector $Z$. 

We define a Markov state vector  at $t$  by $X_t = (F_t, R_t) \in \mathbb{S}_{X} := \mathbb{S}_{F} \times \mathbb{S}_{R}$. The filteration generated by the process $X$ is $\mathcal{F}_{t} := \sigma(\{X_s, s\in[0, t]\})$ and we denote  $X_{0:t} := (X_0, \hdots, X_t)$ as the set of state variables until $t$. The observation, $Z_t = (Y_t^{(j_t)}, \{\hat{Y}_t^{(q_t)}\}_{q_t \in Q_t}) $ at  $t$ is defined using a single physical measurement at $j_t$ and a set of messages $\{\hat{Y}_t^{(q_t)}\}_{q_t \in Q_t}$ for the posterior neighbourhood $Q_{t-1}$ about $j_{t-1}$. The full observational vector is $Z := \{ Z_t, t = 1, 2, \hdots \}$ and  $Z_{0:t} := (Z_0, \hdots, Z_t)$ denotes the measurement record until $t$. 

With these definitions, we can define spaces of probability measures, where each probability measure is a function defined over the space of all possible outcomes for both $X$ and $Z$, satisfying properties associated with the assignment of probabilities. These spaces of probability measures can consist of \textit{random} measures in particle filtering techniques, from which particle filtering methods must converge to the true measure as total data and the total number of particles increase.

The filtering problem is then to compute the specific random measure $\pi_t$ that is the conditional probability of  $X$ given the \ofield{} generated by the observation process $Z_{0:T}$:
\begin{align}
\pi_t &:= \measure{{}}{X_t \in A | \ofieldgen{Z_{0:T}}}, \quad \forall A \in \mathcal{S}_X \\
\pi_t f &= \ex{f(X_t) |  \ofieldgen{Z_{0:T}}} \quad \forall f \in B(\mathbb{S}_X), A \in \mathcal{S}_X \\
\pi_0 & \sim \mathcal{U}(\mathbb{S}_X)
\end{align} In the above, $\pi_t$ is identified with the posterior distribution in Bayesian analysis. The appropriate state space is given by $\mathbb{S}_{X}$ and $\mathcal{S}_X$ is the $\sigma$-algebra generated by $\mathbb{S}_{X}$. Similar comments apply to the observation vector. The term $B(\mathbb{S}_X)$ refers to a space of bounded, $\mathcal{S}_X$-measurable functions which correspond to transformations of the state (e.g. dynamical evolution, measurement models) in the inference procedure.

For the class of particle methods used in our analysis, the transformation of the hidden state $X$ to the observations $Z$ at any $t$ is expressed primarily via the likelihood function. In our case, the likelihood function incorporates both the physical measurement model and the information sharing mechanism. For some constants $\lambda_{1,2} \in [0,1]$,  total likelihood function for each $Z_t=z_t$ is given by:
\begin{align}
g_t^{z_t} &: = g_1(\lambda_1, Y_t^{(j_t)}) g_2(\lambda_2, Q_t) \label{qslam:defn:globallikelihood}
\end{align}
The measurement and map-building noise parameters that govern the level of uncertainty in the system are: $\Sigma_v, \mu_F, \Sigma_F$, where $\Sigma_v$ captures variance of zero mean measurement noise from a quantized sensor Gaussian error model, and $ \mu_F, \Sigma_F$ represent the true error in approximating a continuously varying fields with overlapping Gaussian neighborhoods. The non-negative $\lambda_1, \lambda_2 \in [0,1]$ regulate the extent to which shared information is utilized by the inference procedure.

\subsection{Particle Approximations and Iterative Maximum Likelihood}

The filtering problem, so defined, is challenging as the space of maps and lengthscales is very large. Instead, we use an iterative maximum likelihood approach. This approach assumes that we have access to two data association mechanisms, $h_1$ and $h_2$, to update state variables at $t$ directly using incoming data if the previous state is known. In practical cases, one only has access to estimates of states, and we use  $h_1$ and $h_2$ to update the state estimates of $F_t$ and $R_t$ iteratively within each $t$. To implement these iterative maximum likelihood calculations numerically, we require two different particle sets within our filter. The computational details of these procedures are provide below and they constitute as the iterative maximum likelihood approximation for NMQA.

For $t > 0$, the particle filter has $n_\alpha$ number of particles at the start and the end of an iteration. The set of particles at the beginning and end of each $t$ are called $\alpha$-type particles, indexed by the set of numbers $\{1, 2, \hdots, n_\alpha\}$. For each $\alpha$-particle, we associate a set of $\beta^{(\alpha)}$-particles labeled from $\{1, 2, \hdots, n_\beta \}$. We refer to this as the `$\beta$-layer' for an $\alpha$-particle. A single $\beta^{(\alpha)}$-particle inherits the state $X_t \setminus \{R_t^{(j_t)}\} $ from its $\alpha$-parent; and additionally, inherits a single uniformly distributed sample for $R_t^{(j_t)}$ from the length-scale prior distribution. The total likelihood $g_t^{z_t}$ over the entire particle set is given by the product of the $\alpha$ and $\beta$ particle weights. These conditions yield the following update steps at $t$:
	\begin{enumerate}
	\item Assume $X_{t-1}$ is known. \label{prop:qslam:branching:1}
		\begin{enumerate}
			 \item Weight each $\alpha$-particle according to $ g_1(\lambda_1, Y_t^{(j_t)})$.
			 \item Update $F_t^{(j_t)}$ for each $\alpha$-particle using data via $h_1$. 
		\end{enumerate}
	\item Assume $F_t$ is known. \label{prop:qslam:branching:2}
		\begin{enumerate}
			\item For each $\alpha$-particle, generate $n_\beta$ number of particles of a second type,  called $\beta^{(\alpha)}$-particles.
			\item Weight each $\beta^{(\alpha)}$-particle according $g_2(\lambda_2, Q_t)$.
	\end{enumerate}
	\item Maximize global likelihood $ g_t^{z_t}$:  obtain $n_\alpha$ particles by resampling the full set of $n_\alpha n_\beta$ particles according to the product of their weights \label{prop:qslam:branching:3}
	\item Update $R_t^{(j_t)}$ for each $\alpha$-particle by applying $h_2$ to its $\beta^{(\alpha)}$ layer. \label{prop:qslam:branching:4}
	\item Marginalize over $\beta^{(\alpha)}$ layer yielding $n_\alpha$ number of $\alpha$-particles with uniform weights.  \label{prop:qslam:branching:5}
	\end{enumerate} 
	
	Collectively, these steps form the branching mechanism for particles in the NMQA filter. The word `branching' refers to the concept of replacing the set of particles at the start of the $t$ (parents) with a new set of particles (off-spring) at the end of the iteration, after measurement data is received. The updates are performed such that the global likelihood $g_t^{z_t}$ is maximized. 

The data association mechanisms $h_1$ and $h_2$ are written as a set of recursive equations which conform to both a Markov description of our state space and an iterative, numerical inference procedure. In the equations below,  $h_1(\lambda_1, Z_t)$ updates the state variable $F_t^{(j)}$  given $X_{t-1}$ (omitting $t$ sub-script on $j$ for ease of reading) and $h_2$ updates the state variable $R_t^{(j)}$ given $F_t$. 
	\begin{align}
	F_t^{(j)} &:= \begin{cases}
	F_{t-1}^{(j)}, \quad \tau_t^{(j)}, \varphi_t^{(j)} = 0 \\
	h_1(\lambda_1, Z_t), \quad \mathrm{otherwise}   \\
	\end{cases} \\
	R_t^{(j)} &:= h_2(F_t) \\
	h_1(\lambda_1, Z_t)& := \cos^{-1} (2P_t^{(j)}(\lambda_1, Z_t) - 1) \\
	h_2(F_t)&= \ex{\{R_t^{(j), n}\}_{n=1}^{n_\beta} | F_t} \\
	P_t^{(j)}(\lambda_1, Z_t) &:= \begin{cases}
	\left( 1- \frac{\lambda_1^{\tau_t^{(j)}}}{2}\right)\kappa_{t}^{(j)} + \left(\frac{\lambda_1^{\tau_t^{(j)}}}{2}\right)\gamma_{t}^{(j)}, \\ \text{for } \tau_t^{(j)}, \varphi_t^{(j)} \neq 0. \\
	\kappa_{t}^{(j)}, \text{ for } \tau_t^{(j)}\neq 0, \varphi_t^{(j)} = 0. \\
	\gamma_{t}^{(j)}, \text{ for } \tau_t^{(j)} = 0, \varphi_t^{(j)} \neq 0. \\
	\end{cases}\\
	\kappa_{t}^{(j)} &:= \frac{\tau_{t-1}^{(j)}}{\tau_{t}^{(j)}}  \kappa_{t-1}^{(j)} + \frac{1}{\tau_t^{(j)}} Y_t^{(j_t)}I_{(j_t = j)}, \nonumber \\
	& \tau_t^{(j)} \neq 0, 	\kappa_{0}^{(j)} =0 \\
	\gamma_{t}^{(j)} &:= \frac{\varphi_{t-1}^{(j)} }{\varphi_{t}^{(j)}} \gamma_{t-1}^{(j)} + \frac{1}{\varphi_t^{(j)}} \hat{Y}_t^{(q_t)}I_{(q_t= j, q_t \in Q_t)} , \nonumber \\ & \varphi_t^{(j)} \neq 0, \gamma_{0}^{(j)} =0\\
	\tau_t^{(j)} &:= \tau_{t-1}^{(j)} + I_{(j_t = j)}, \quad \tau_0^{(j)} =0 \\
	\varphi_t^{(j)} &:= \varphi_{t-1}^{(j)} + I_{(q_t= j, q_t \in Q_t)}, \quad \varphi_0^{(j)} =0 
	\end{align} Here, $I_{X}$  is the indicator function equalling unity if $X$, and zero if not $X$. We see that $P_t^{(j)}$ is just an empirical Born probability estimate obtained by averaging the binary measurement data obtained at a single qubit. This data consists of averaging over $\tau_t^{(j)}$ physical measurements at qubit $j$,  and $\varphi_t^{(j)}$ shared messages induced by measuring the neighboring qubits of $j$. The role of shared messages is reduced as the number of physical measurements, $\tau_t^{(j)}$, increases, at a rate governed by the choice of $\lambda_1$. The data association mechanism $h_2$ locally updates the state variable $R_t^{(j)}$ as the expectation over resampled $\beta$-particles given the state at $F_t$. The dependence of  $h_2$ on $F_t$ is implicit, as $\beta$-particles are weighted via the $g_2(\lambda_2, Q_t)$ likelihood function for which it is assumed $F_t$ is known.

With the above computations, NMQA can be numerically implemented in software. Python-packages for the NMQA algorithm, data generation scripts and simulation analysis are accessible via http://github.com/qcl-sydney/quantum\_slam.

\section{Special Functions}

The following notation refers to special functions:

\begin{align}
    \mathcal{Q}(x) &:= \text{Bernoulli trial with win probability $x$.} \\
    \delta(x) &:= \text{Kronecker-delta function in $x$.} \\
    \mathrm{erf}(x) &:= \text{Error function in $x$.}\\
    I_X &:= \text{Indicator function;  $1$ if $X$, else $0$.}
\end{align}